\documentclass[epj]{svjour}
\usepackage{graphics}
\usepackage{amsmath}
\usepackage{amsfonts}
\usepackage{amssymb}
\usepackage[normalem]{ulem}
\def\half{\tfrac{1}{2}}
\begin{document}

\title{{Non-equilibrium concentration fluctuations in binary liquids with realistic boundary conditions }}
\titlerunning{Non-equilibrium fluctuations in binary liquids\dots}

\author{J. M. Ortiz de Z\'arate\inst{1} \and T. R. Kirkpatrick \inst{2,3} \and J. V. Sengers\inst{2}% etc
}                     % Do not remove
\institute{Departamento de F\'{\i}sica Aplicada I, Facultad de F\'{\i}sica, Universidad Complutense, 28040 Madrid, Spain \and
Institute for Physical Science and Technology, University of Maryland, College Park, Maryland 20742, USA \and
Department of Physics, University of Maryland, College Park, Maryland 20742, USA}
\date{Received: date / Revised version: date}

\date{\today}% It is always \today, today, but you may specify any date with \date.

\abstract{%
Because of the spatially long-ranged nature of spontaneous fluctuations in thermal non-equilibrium systems, they are affected by boundary conditions for the fluctuating hydrodynamic variables. In this paper we consider a liquid mixture between two rigid and impervious plates with a stationary concentration gradient resulting from a temperature gradient through the Soret effect. For liquid mixtures with large Lewis and Schmidt numbers, we are able to obtain explicit analytical expressions for the intensity of the non-equilibrium concentration fluctuations as a function of the frequency $\omega$ and the wave number $q$ of the fluctuations.  In addition we elucidate the spatial dependence of the intensity of the non-equilibrium fluctuations responsible for a non-equilibrium Casimir effect.
}%

 % end of PACS codes

\maketitle

\section{Introduction}

It has been well established, both theoretically and experimentally, that thermal fluctuations in fluids in the presence of a temperature gradient and/or a concentration gradient are spectacularly long range~\cite{KirkpatrickEtAl,DorfmanKirkpatrickSengers,BOOK}. Specifically, the intensity of the temperature and concentration fluctuations in non-equilibrium (NE) fluids varies with the wave number $q$  of the fluctuations as $q^{-4}$ which means that in real space the correlations extend over the size $L$ of the system~\cite{Physica}. As a consequence, the spatial spectrum of these NE fluctuations is strongly affected by finite-size effects due to the presence of boundaries, especially for small wave numbers accessible in shadowgraph experiments. Hence, for the interpretation of NE shadowgraph experiments~\cite{OOSA03,VailatiEtAl2,CroccoloEtAl,TakacsEtAl,TakacsEtAl2,VailatiEtAl,OprisanPayne,CerbinoEtAl} and for the validation of computational studies~\cite{BalboaEtAl,DelongEtAl,BalakrishnanEtAl,DelongEtAl2,DonevEtAl3}  of NE fluctuations, an assessment of the impact of the boundary conditions of the fluctuating hydrodynamic variables on the NE fluctuations is necessary.

A procedure for solving the fluctuating hydrodynamics equations to obtain the intensity of the NE concentrations fluctuations has been developed by two of us, but with artificial boundary conditions for the fluctuations at the walls adopted for mathematical convenience~\cite{Mexico}. For the case of realistic boundary conditions, we subsequently obtained an approximate solution in terms of a Galerkin approximation~\cite{miIMT6}.

In liquid mixtures there are two diffusion modes that are linear combinations of heat and mass diffusion and a viscous mode~\cite{Wood,Mixtures3}. Important parameters are the Lewis number, that is the ratio of the thermal diffusivity $D_T$ and the mass diffusion coefficient $D$, and the Schmidt number, that is the ration of the kinematic viscosity $\nu$ and $D$. In liquid mixtures both Lewis and Schmidt numbers are commonly larger than unity, implying that  temperature and viscous fluctuations decay much faster than concentrations fluctuations. Hence, in dealing with liquids one often adopts an approximation of large Lewis and Schmidt numbers, $Le\gg1$ and $Sc\gg1$. This approach is particularly convenient when the focus is on concentration fluctuations at time scales when temperature and viscous fluctuations are fully decayed~\cite{BOOK,VelardeSchechter}.
For $Le\gg1$  and $Sc\gg1$ we were able to obtain an exact expression for the autocorrelation function of the intensity of the NE concentration fluctuations in the presence of realistic boundary conditions, but that required a tedious numerical evaluation of the decay rates (eigenvalues) of the modes of the hydrodynamic operator, yielding some analytical results only in the long-wavelength limit~\cite{miPRE2}. Subsequently, the procedure was extended to study the dynamics of the NE concentration fluctuations and good agreement with experimental measurements was obtained, although again a tedious numerical evaluation of the appropriate eigenvalues was required~\cite{miDynamics14}.

NE concentration fluctuations are not only affected by finite-size effects, but also by the presence of gravity.  In this paper, as in some earlier work~\cite{Mexico}, we focus our attention specifically on the finite-size effects. We have been able to obtain relatively compact analytic expressions for both the dynamic and static autocorrelation function of the NE concentration fluctuations that do require only minimal numerical work.  Our results are not only relevant to past and future experimental studies of NE concentration fluctuations at low gravity~\cite{VailatiEtAl,MialdunEtAl}, but also for certain aspects of earth-bound non-equilibrium phenomena, like a NE Casimir effect~\cite{miPRL2,miPRE2014,miCasimirBin}, as well as for the interpretation of numerical simulations of non-equilibrium fluctuating hydrodynamics in mixtures~\cite{DelongEtAl2,DonevEtAl3,DonevEtAl2}.

We shall proceed as follows. The equations for the appropriate fluctuating variables and boundary conditions are specified in Section~\ref{S2}. The concentration gradient induces a coupling between concentration fluctuations and wall-normal velocity fluctuations along the direction of the concentration gradient~\cite{BOOK,LawNieuwoudt,SegreEtAlPRE}. The equation for these velocity fluctuations is solved in Section~\ref{S3}. Inserting this solution in the equation for the concentration fluctuations we obtain the correlation functions for the NE concentration fluctuations in Section~\ref{S4}. In Section~\ref{S5} we analyze the spatial dependence of the intensity of the NE concentration fluctuations needed for an understanding of Casimir forces induced by the NE concentration fluctuations~\cite{miCasimirBin}. Our results are summarized and discussed in Section~\ref{S6}.

\section{Fluctuating variables and boundary conditions\label{S2}}

We consider a layer of a binary fluid mixture bounded by two plane-parallel walls separated by a distance $L$ and located at $z=\pm\half L$. The plates are maintained at different temperatures, so that a stationary temperature gradient $\boldsymbol{\nabla}T=(\nabla{T})~\hat{\vec{z}}$ exists, with $\hat{\vec{z}}$ the unit vector in the wall-normal direction. As a consequence of the Soret effect, a stationary concentration gradient $\boldsymbol{\nabla}c=(\nabla{c})~\hat{\vec{z}}$ will also appear, where $c$ is the concentration in mass fraction of one of the components of the mixture. We shall further consider that all relevant thermophysical properties of the mixture are constant (do not depend on temperature or concentration) so that both $\nabla{T}$ and $\nabla{c}$ are uniform. This physical situations is usually referred to as the Rayleigh-B\'enard problem for a binary mixture~\cite{VelardeSchechter,SchechterEtAl}. In the absence of gravity, the stationary state described above is stable. Finally, consistent with the $Le\gg1$ approximation, we assume the separation ratio of the mixture $\psi=-\beta\nabla{c}/\alpha\nabla{T}$ to be positive~\cite{VelardeSchechter,LectureNotes} (the most common case). Here $\alpha=-1/\rho~(\partial\rho/\partial{T})_{c,p}$ is the thermal expansion coefficient and $\beta=1/\rho~(\partial\rho/\partial{c})_{T,p}$ the solutal expansion coefficient, with $\rho$ the mass density of the mixture.

Fluctuating hydrodynamics, originally developed for dealing with spontaneous thermodynamic fluctuations in equilibrium~\cite{LandauLifshitz,FoxUhlenbeck1,Foch1}, can also account for fluctuations around the stationary gradient specified above~\cite{BOOK,LawNieuwoudt}. In the most general case, as in equilibrium~\cite{Foch1}, there will be coupled pressure, temperature, concentration and velocity fluctuations, so that some simplifications are desirable. If $\delta\vec{v}$ are the velocity fluctuations, one first assumes, as usual for dense fluids, that they are divergence-free: $\boldsymbol{\nabla}\cdot\delta\vec{v}$ (incompressible flow). This assumption allows us to neglect the sound modes~\cite{LawNieuwoudt}. Next, we adopt here a large Lewis number approximation ($Le\gg1$) that decouples temperature and concentration fluctuations. Hence, one has to consider only the coupling of concentration and wall-normal velocity fluctuations~\cite{Mexico,miPRE2,LectureNotes}. The $Le\gg1$ approximation was first proposed to simplify the linear stability analysis of the Rayleigh-B\'enard problem in binary fluids~\cite{VelardeSchechter,SchechterEtAl}, demonstrating its validity for positive separation ratios. Subsequently, it was used for simplifying the associated fluctuating-hydrodynamics problem~\cite{Mexico,miPRE2,LectureNotes}. It has been successfully employed in numerical simulations~\cite{BalakrishnanEtAl}, for reproducing light-scattering experiments in binary mixtures~\cite{miDynamics14} and, more recently, in ternary mixtures~\cite{DonevEtAl3,miEPJAnglet}. We note that for the most common liquid mixtures, $Le$ is of the order $10-10^3$.

The fluctuating hydrodynamics equations for our problem have been presented in some previous publications, to which we refer for details and physical background~\cite{Mexico,miPRE2,LectureNotes}. For our purpose here we find it convenient to use the working equations of Ref.~\cite{miPRE2} in dimensionless form. Hence, we introduce dimensionless space, time and concentration variables by:
\begin{align}\label{E03}
\tilde{r}&=r/L,&\tilde{t}&=\frac{tD}{L^2},&{\delta\widetilde{c}}&=\frac{\delta{c}}{\nabla{c}~L}.
\end{align}
In terms of these dimensionless variables, and in the absence of gravity, the fluctuating hydrodynamics equations for a binary fluid mixture in the $Le\gg1$ limit become~\cite{miPRE2}:
\begin{subequations}\label{E1}
\begin{align}
\frac{1}{Sc}\partial_t(\nabla^2\delta{v}_z)&=\nabla^4\delta{v}_z-F_1(\vec{r},t)\label{E1A}\\
\partial_t\delta{c}&=\nabla^2\delta{c}-\delta{v}_z+ F_2(\vec{r},t)
\end{align}
\end{subequations}
where $Sc=\nu/D$ is the Schmidt number and
\begin{equation}\label{ERF}
\begin{split}
F_1(\vec{r},t)&=-\frac{L^2}{\rho\nu{D}}\left\{\boldsymbol{\nabla}\times\boldsymbol{\nabla}\times(\boldsymbol{\nabla}\cdot\delta\vec{\Pi})\right\}_z,\\
F_2(\vec{r},t)&=-\frac{L}{\rho D\nabla{c}}~ \boldsymbol{\nabla}\cdot\delta\vec{J}.
\end{split}
\end{equation}
We note that in Eqs.~\eqref{E1}-\eqref{ERF}, for simplicity, tildes in dimensionless variables have been suppressed. From here on, everything will be in terms of dimensionless variables, except where explicitly the opposite is stated.

As in our previous work~\cite{miPRE2}, to further simplify Eqs.~\eqref{E1} we also assume the Schmidt number to be large, $Sc\gg1$. Hence, in practice, the left-hand side (LHS) of Eq.~\eqref{E1A} will be taken as zero. As $Le\gg1$, $Sc\gg1$ is also a good approximation for typical liquid mixtures. This approximation amounts to assuming that viscous fluctuations decay very fast in the time scale set by the diffusion time, see Eq.~\eqref{E03}.

Equations~\eqref{ERF} define dimensionless random forces expressed as spatial derivatives of the two fluctuating dissipative fluxes: a random stress tensor $\delta\vec{\Pi}(\vec{r},t)$ and a random diffusion flux $\delta\vec{J}(\vec{r},t)$. We note that, in fluctuating hydrodynamics~\cite{BOOK,LawNieuwoudt,LandauLifshitz,FoxUhlenbeck1,Foch1}, the linear phenomenological relations, used to 'close' the balance laws and to obtain the hydrodynamic equations, need to be supplemented with fluctuating dissipative fluxes. In our case, a $\delta\vec{\Pi}$ is added to Newton's viscosity law, and a $\delta\vec{J}$ is added to Fick's diffusion law. The statistical properties of these fluctuating dissipative fluxes are given, in terms of the dissipation matrix, by the so-called Fluctuation Dissipation Theorem (FDT). Since the random diffusion flux contributes only to equilibrium fluctuations, in what follows only the FDT for the random stress tensor will be needed, which for the case of incompressible flow and in terms of dimensional variables, it explicitly reads~\cite{BOOK,miPRE2}:
\begin{multline}\label{FDT}
\langle\delta\Pi_{ij}(\vec{r},t)~\delta\Pi_{kl}^*(\vec{r}^\prime,t^\prime)\rangle=2k_\text{B}T\eta\\ \times (\delta_{ik}\delta_{jl}+\delta_{il}\delta_{jk})~\delta(t-t^\prime)~\delta(\vec{r}-\vec{r}^\prime).
\end{multline}
where $k_\text{B}$ is Boltzmann's constant and $\eta$ the shear viscosity. As previous investigators~\cite{SchmitzCohen1,SchmitzCohen2} we assume that a local version of the FDT continues to be valid in non equilibrium, but we neglect the effects of nonhomogeneous noise so that, in practice, temperature $T$ in Eq.~\eqref{FDT} is identified with the average temperature in the layer.

Realistic boundary conditions are no-slip for the velocity fluctuations and impervious wall for the mass flow. For an incompressible fluid, and in terms of dimensionless variables, they are~\cite{Chandra}:
\begin{align}\label{E2}
0&=\delta{v}_z=\partial_z\delta{v}_z, & 0&=\partial_z\delta{c}, &\text{at},~&z=\pm\half.
\end{align}
In the second of the boundary conditions above the limit $Le\gg1$ is implicit, meaning that temperature fluctuations are neglected and consequently the thermodiffusion contribution to the mass flow.

Next, to solve Eqs.~\eqref{E1} with the boundary conditions~\eqref{E2}, as usual~\cite{BOOK,miPRE2,SchmitzCohen1,SchmitzCohen2}, one applies Fourier transforms in time and in the $XY$-plane, parallel to the walls. Hence, in matrix form and for $Sc\gg1$, Eqs.~\eqref{E1} become:
\begin{equation}\label{E3}
\begin{bmatrix}
(\partial_z^2-q^2)^2&0\\
1&\mathrm{i}\omega+q^2-\partial_z^2
\end{bmatrix}
\begin{bmatrix}
\delta{v}_z\\
\delta{c}
\end{bmatrix}
=\begin{bmatrix}
F_1(\omega,\vec{q},z)\\
F_2(\omega,\vec{q},z)
\end{bmatrix},
\end{equation}
where $\omega$ is the frequency of the fluctuations and $\vec{q}$ is the component of the wave vector of the fluctuations in the horizontal plane (with corresponding wave number $q$). The random forces $F_i(\omega,\vec{q},z)$ in Eq.~\eqref{E3} are the Fourier transforms of the random forces in Eq.~\eqref{ERF}. To solve Eq.~\eqref{E3} we first separate its right-hand side (RHS) as~\cite{miEPJAnglet}:
\begin{equation}
\begin{bmatrix}
F_1(\omega,\vec{q},z)\\
F_2(\omega,\vec{q},z)
\end{bmatrix}=
\begin{bmatrix}
0\\
F_2(\omega,\vec{q},z)
\end{bmatrix}
+\begin{bmatrix}
F_1(\omega,\vec{q},z)\\
0
\end{bmatrix}.
\end{equation}
This splits the solution of Eq.~\eqref{E3} additively into two parts: $\delta{c}=\delta{c}^\text{E}+\delta{c}^\text{NE}$ that we distinguish with superscripts 'E' and 'NE'. It is interesting to note that the random stress tensor $\delta\vec{\Pi}$ and the random diffusion flux $\delta\vec{J}$ are uncorrelated, so that the random forces $F_1$ and $F_2$ in the RHS of Eq.~\eqref{E3} are uncorrelated too, see Eq.~\eqref{ERF}. As a consequence, the autocorrelation function of the concentration fluctuations splits additively also into 'E' and 'NE' components. Furthermore the 'E' component of the autocorrelation is the same as for the system in equilibrium, \emph{i.e.}, for $\nabla{c}=0$ (or the 1 were not present in the first column of the hydrodynamic matrix of Eq.~\eqref{E3}). From now on we concentrate on the 'NE' contribution to the concentration fluctuations to be determined from Eq.~\eqref{E3}, namely:
\begin{equation}\label{E5}
\begin{bmatrix}
(\partial_z^2-q^2)^2&0\\
1&\mathrm{i}\omega+q^2-\partial_z^2
\end{bmatrix}
\begin{bmatrix}
\delta{v}_z\\
\delta{c}
\end{bmatrix}
=\begin{bmatrix}
F_1(\omega,\vec{q},z)\\
0
\end{bmatrix},
\end{equation}
where, for simplicity, we dropped superscript 'NE'. Of course, Eq.~\eqref{E5} must be solved subjected to the boundary conditions~\eqref{E2} in the wall-normal $z$-variable.

\section{Wall-normal velocity fluctuations\label{S3}}

We start by solving the velocity equation, the first of Eqs.~\eqref{E5}. As in previous works, the solution is conveniently expressed in a series of eigenfunctions~\cite{Physica,miIMT6,miPRE2,SchmitzCohen2}
\begin{equation}\label{E12}
\delta{v}_z(z)=\sum_{N=1}^\infty C_N~W_N(z),
\end{equation}
where the hydrodynamic modes $W_N(z)$ are the solutions of the eigenvalue problem
\begin{equation}\label{E13}
(\partial_z^2-q^2)^2 W_N(z)=\Gamma_N^2~W_N(z),
\end{equation}
subject to the boundary conditions
\begin{align}\label{E14}
W_N(z)&=\partial_z~W_N(z)=0,&\text{at~}z&=\pm\half.
\end{align}
In Eqs.~\eqref{E13}-\eqref{E14} we have used the fact that the eigenproblem~\eqref{E13} has an infinite numerable set of solutions $W_N(z)$ with distinct eigenvalues $\Gamma_N^2$ that are real and positive numbers. Although we have explicitly calculated the solution to Eqs.~\eqref{E13}-\eqref{E14} in terms of hyperbolic functions, including the numerical solution, for each $q$, of an algebraic equation to obtain $\Gamma_N^2$, we shall not present it here. It is very technical and, as clarified below, we only need the trace of the differential operator in the LHS of Eq.~\eqref{E13} with boundary conditions~\eqref{E14}. As happens in many problems in physics of fluids~\cite{BamiehDahleh,miORR}, the trace of the hydrodynamic operator can be computed without detailed discussion of the hydrodynamic modes.

Indeed, independently of the details of $W_N(z)$, it follows that:
\begin{multline}
\int_{-1/2}^{1/2}\hspace*{-12pt}dz~W_M(z)\left[(\partial_z^2-q^2)^2W_N(z)\right]=\\
\int_{-1/2}^{1/2}\hspace*{-12pt}dz\left[(\partial_z^2-q^2)^2W_M(z)\right]~W_N(z),
\end{multline}
obtained upon integration by parts and use of the boundary conditions~\eqref{E14}. As a consequence, the $W_N(z)$ form an ortogonal set, namely
\begin{equation}\label{E16}
\int_{-1/2}^{1/2}\hspace*{-12pt}dz~W_M(z)~W_N(z)=D_{N}~\delta_{NM}.
\end{equation}
Hence, from Eqs.~\eqref{E13} and~\eqref{E16}, the coefficients $C_N$ in the series expansion~\eqref{E12} can be readily evaluated, resulting in:
\begin{equation}\label{E15}
C_N=\frac{F_{1,N}(\omega,\vec{q})}{D_N(q)~\Gamma_N^2(q)},
\end{equation}
with
\begin{equation}
F_{1,N}(\omega,\vec{q})=\int_{-1/2}^{1/2}\hspace*{-12pt}dz~W_N(q,z)~F_1(\omega,\vec{q},z).
\end{equation}
This procedure solves the first of Eqs.~\eqref{E5} with the appropriate boundary conditions.

However, we are interested only in the two-point correlation function $\langle\delta{v}_z(z)~\delta{v}_z^*(z^\prime)\rangle$. By using Eq.~\eqref{E12} it can be expressed as a double series of hydrodynamic modes $W_N(z)$. From Eq.~\eqref{E15} it follows that the coefficients of this double series are related to the correlation matrix $\langle{F}_{1,N}(\omega,\vec{q})~{F}_{1,M}^*(\omega^\prime,\vec{q}^\prime)\rangle$.
This correlation matrix, in turn, can be calculated from the definition of $F_1(\vec{r},t)$, Eq.~\eqref{ERF}, and the FDT for the random stress $\delta\vec{\Pi}(\vec{r},t)$, Eq.~\eqref{FDT}. This last calculation has been already presented in detail previously~\cite{BOOK,miORR}. Here we simply quote the final result:
\begin{multline}\label{E19}
\langle{F}_{1,N}(\omega,\vec{q})~{F}_{1,M}^*(\omega^\prime,\vec{q}^\prime)\rangle=F_{NM}(q) \\ \times (2\pi)^3~\delta(\omega-\omega^\prime)~\delta(\vec{q}-\vec{q}^\prime),
\end{multline}
with
\begin{align}
F_{NM}(q)&=2\tilde{F}q^2 \iint_{-1/2}^{1/2}\hspace*{-12pt}dzdz^\prime~[\mathcal{D}_z^2~W_N(q,z)]~W^*_M(q^\prime,z^\prime),\notag\\
&=2\tilde{F}q^2 \Gamma_N^2(q)~D_N(q)~\delta_{NM}.\label{E20}
\end{align}
where, in the second line of Eq.~\eqref{E20}, we used Eqs.~\eqref{E13} and~\eqref{E16}. In Eq.~\eqref{E20} $\mathcal{D}_z^2=(\partial_z-q^2)^2$, while the symbol $\tilde{F}$ denotes a dimensionless prefactor
\begin{equation}\label{E19B}
\tilde{F}= \frac{k_\text{B} T}{\rho\nu DL}.
\end{equation}
Now combining all the information above we are able to obtain an explicit expression for the wall-normal velocity autocorrelation function. In view of Eq.~\eqref{E19} one has:
\begin{multline}\label{E21}
\langle\delta{v}_z(\omega,\vec{q},z)~\delta{v}_z^*(\omega^\prime,\vec{q}^\prime,z^\prime)\rangle= V(q,z,z^\prime)\\
\times(2\pi)^3~\delta(\omega-\omega^\prime)~\delta(\vec{q}-\vec{q}^\prime),
\end{multline}
with
\begin{equation}\label{E22}
V(q,z,z^\prime) = \sum_{N=0}^\infty \frac{2\tilde{F} q^2}{D_N(q)~\Gamma_N^2(q)}W_N(q,z)~W_N(q,z^\prime),
\end{equation}
where we use the fact that the eigenvalues and eigenfunctions are real and positive. It is interesting to note that the autocorrelation $V(q,z,z^\prime)$ in Eq.~\eqref{E21} does not depend explicitly on the frequency $\omega$. This is a consequence of the $Sc\gg1$ approximation, which neglects the inertial term in the starting equations, that is, the LHS of Eq.~\eqref{E1A} is assumed to be zero~\cite{miPRE2,miDynamics14}. If the temporal derivative of the velocity fluctuations were retained in the LHS of Eq.~\eqref{E1A}~\cite{BalakrishnanEtAl}, the the wall-normal velocity autocorrelation would depend explicitly on $\omega$.

As in previous publications~\cite{BamiehDahleh,miORR}, the sum of the series~\eqref{E22} can be performed analytically. Indeed, applying the operator $(\partial_z^2-q^2)^2$ at the two members of Eq.~\eqref{E22} one has:
\begin{equation}\label{E23}
(\partial_z^2-q^2)^2~V(q,z,z^\prime) = 2 \tilde{F} q^2~\delta(z-z^\prime).
\end{equation}
Notice that, since $\delta(z-z^\prime)$ as a function of $z$ satisfies the boundary conditions~\eqref{E14}, it can be expanded in a series of eigenfunctions $W_N(z)$, namely
\begin{equation}\label{E24}
\delta(z-z^\prime) = \sum_{N=0}^\infty \frac{1}{D_N(q)}W_N(q,z)~W_N(q,z^\prime),
\end{equation}
independent of $q$, from which Eq.~\eqref{E23} follows. Hence, if one solves Eq.~\eqref{E23} subject to the boundary conditions
\begin{align}\label{E25}
V(q,z,z^\prime)&=\partial_z~V(q,z,z^\prime)=0,&\text{at~}z&=\pm\half,
\end{align}
which follows from Eq.~\eqref{E14}, we obtain an analytic expression for $V(q,z,z^\prime)$ presented in Appendix~\ref{A2} (see Eq.~\eqref{E07}).

\section{Non-equilibrium concentration fluctuations\label{S4}}

To solve the second of Eqs.~\eqref{E5} for the concentration fluctuations with the appropriate boundary conditions, we follow an equivalent approach to the one in the previous section for the wall-normal velocity fluctuations. That is, we expand $\delta{c}$ in a series of eigenfunctions of the differential operator $\partial_z^2-q^2$ subject to the boundary conditions. In this case this approach is quite simple since the eigenfunctions are trigonometric functions. Hence, one has:
\begin{multline}\label{E29B}
\delta{c}(\omega,q,z) = \frac{-A_0}{\mathrm{i}\omega+q^2}-2\sum_{N=1}^\infty \frac{A_N \cos(2N\pi z)}{\mathrm{i}\omega+q^2+4N^2\pi^2}\\
-2\sum_{N=0}^\infty \frac{B_N \sin[(2N+1)\pi z]}{\mathrm{i}\omega+q^2+(2N+1)^2\pi^2}
\end{multline}
with
\begin{equation}
\begin{split}
A_N(\omega,q)&=\int_{-1/2}^{1/2} dz~\delta{v}_z(\omega,q,z)~\cos(2N\pi z),\\
B_N(\omega,q)&=\int_{-1/2}^{1/2} dz~\delta{v}_z(\omega,q,z)~\sin[(2N+1)\pi z].
\end{split}
\end{equation}
From Eq.~\eqref{E29B} we see that the autocorrelation function $\langle\delta{c}(\omega,\vec{q},z) \delta{c}^*(\omega^\prime,\vec{q}^\prime,z^\prime)\rangle$ of the NE concentration fluctuations is directly related to $\langle\delta{v}_z(\omega,\vec{q},z)~\delta{v}_z^*(\omega^\prime,\vec{q}^\prime,z^\prime)\rangle$, the autocorrelation function of the wall-normal velocity fluctuations discussed in the previous section. In view of Eq.~\eqref{E21}, the result can be cast in the form
\begin{multline}\label{E26}
\langle\delta{c}(\omega,\vec{q},z)~\delta{c}^*(\omega^\prime,\vec{q}^\prime,z^\prime)\rangle= S(\omega,q,z,z^\prime)\\
\times(2\pi)^3~\delta(\omega-\omega^\prime)~\delta(\vec{q}-\vec{q}^\prime),
\end{multline}
where $S(\omega,q,z,z^\prime)$ is a double series of trigonometric functions. For generic $z$ and $z^\prime$, the explicit expression for $S(\omega,q,z,z^\prime)$ is quite lengthy, so that we present it explicitly in Appendix~\ref{A3}, see in particular Eq.~\eqref{E34}. Fortunately, for practical applications, one does not need the general $S(\omega,q,z,z^\prime)$, but some particular values and/or integrals that allow for more compact expressions.

In light-scattering~\cite{BOOK,Mixtures3,SegreEtAlPRE} or shadowgraph~\cite{miDynamics14,TrainoffCannell} experiments one commonly studies the NE fluctuations with the wave vector in the horizontal direction. The structure factor $S(\omega,q)$ that is measured in such experiments is given by a double integration over the thickness of the layer of the $S(\omega,q,z,z^\prime)$ of Eq.~\eqref{E26}, namely:
\begin{equation}\label{E29}
S(\omega,q)=\int_{-1/2}^{1/2}\hspace*{-12pt}dz \int_{-1/2}^{1/2}\hspace*{-12pt}dz^\prime ~S(\omega,q,z,z^\prime).
\end{equation}
Integrating Eq.~\eqref{E34} for $S(\omega,q,z,z^\prime)$, one readily observes that only the first term of the series (that does not depend on $z$ or $z^\prime$) gives a non-zero contribution to $S(\omega,q)$. Hence, one simply has:
\begin{equation}\label{E30}
{S(\omega,q)=\frac{2\tilde{F}}{\omega^2+q^4}\left[\frac{1}{q^2} + \frac{4(1-\cosh{q})}{q^3(q+\sinh{q})} \right].}
\end{equation}
Applying double inverse Fourier transforms to Eq.~\eqref{E26}, and taking into account Eq.~\eqref{E29}, one obtains the intensity $S(q)$ of the equal-time NE concentration fluctuations observed in experiments:
\begin{multline}
\int_{-1/2}^{1/2}\hspace*{-12pt}dz \int_{-1/2}^{1/2}\hspace*{-12pt}dz^\prime~\langle\delta{c}(t,\vec{q},z)~\delta{c}^*(t,\vec{q}^\prime,z^\prime)\rangle\\
=S(q)~(2\pi)^2~\delta(\vec{q}-\vec{q}^\prime).
\end{multline}
Integration of Eq.~\eqref{E30} over the frequency $\omega$ yields a compact exact analytic expression for $S(q)$
\begin{equation}\label{E31}
\begin{split}
S(q)&=\frac{1}{2\pi}\int_{-\infty}^\infty d\omega~ S(\omega,q)\\
&=\tilde{F}\left[\frac{1}{q^4} + \frac{4(1-\cosh{q})}{q^5(q+\sinh{q})} \right].
\end{split}
\end{equation}
From Eq.~\eqref{E31} we obtain at large $q\gg1$
\begin{equation}
S(q) \simeq \frac{\tilde{F}}{q^4}-\frac{4\tilde{F}}{q^5}+ \cdots
\end{equation}
whose first term contains the typical $S(q)\simeq q^{-4}$ behavior of NE temperature and concentration fluctuations discussed in many publications~\cite{KirkpatrickEtAl,LawNieuwoudt,SegreSengers}. For this particular problem of NE concentration fluctuations induced by the Soret effect, the $q^{-4}$ behavior was first found by Law and Nieuwoudt~\cite{LawNieuwoudt} without gravity and by Segr\'e and Sengers~\cite{SegreSengers} with gravity. In the limit $q\to0$, $S(q)$ from Eq.~\eqref{E31} reaches a finite limit, namely
\begin{equation}\label{E31A}
S(q) \simeq \frac{\tilde{F}}{720}-\frac{\tilde{F}q^2}{15120} + \mathcal{O}(q^4)
\end{equation}
in agreement with Ref.~\cite{miPRE2} for a solutal Rayleigh number equal to zero (no gravity).

All results presented so far are in terms of the dimensionless variables defined in Eq.~\eqref{E03}. In terms of dimensional variables our primary result, the NE structure factor given by Eq.~\eqref{E30}, becomes:
\begin{equation}\label{E31B}
{S(\omega,q)=\frac{k_\text{B} T}{\rho} \frac{(\nabla{c})^2}{\nu D q^4} \frac{2Dq^2}{\omega^2+D^2q^4}\left[1+ \frac{4(1-\cosh{\tilde{q}})}{\tilde{q}(\tilde{q}+\sinh{\tilde{q}})} \right].}
\end{equation}
where $\tilde{q}=qL$. In obtaining Eq.~\eqref{E31B}, Eq.~\eqref{E03} for dimensionless concentration and Eq~\eqref{E19B} for $\tilde{F}$, were used.

It is interesting to observe, either in Eq.~\eqref{E30} or in Eq.~\eqref{E31B}, that in the absence of gravity the NE concentration fluctuations have a single decay time, even when boundary conditions are included in the calculation. This is in contrast with the situation when gravity is not neglected, for which a series of exponentials is obtained for the NE time correlation function~\cite{miDynamics14}. The physical reason behind this difference is that gravity affects the decay time of NE concentration fluctuations, not only the amplitude, as first discussed by Segr{\`e} and Sengers~\cite{SegreSengers}. That in microgravity NE concentration fluctuations have a single decay time, even at very small $q$, as predicted by Eqs.~\eqref{E30} or~\eqref{E31B}, has indeed been observed experimentally~\cite{TakacsEtAl2}.

The exact result of Eq.~\eqref{E31} supersedes a Galerkin approximation proposed some years ago~\cite{miIMT6} which, in the limit of zero gravity, was used for the analysis of microgravity experiments~\cite{VailatiEtAl,BalboaEtAl}. Comparison of the two expressions shows that both share the same asymptotic limit for large $q$, while the Galerkin approximation~\cite{miIMT6} overestimates the exact small $q$ limit of Eq.~\eqref{E31A} by $\simeq30\%$.

\section{Correlations relevant for the NE Casimir effect\label{S5}}

Because of their spatially long-ranged nature, it has been recently proposed that NE fluctuations will induce forces (Casimir forces) somewhat similar to the Casimir forces appearing at equilibrium critical points. To evaluate these NE Casimir forces~\cite{miPRL2,miPRE2014,miCasimirBin} one expands the pressure up to second order in the fluctuating fields, so that the resulting fluctuation-induced force becomes proportional to the mean square intensity of the NE fluctuations in real space. For the case of a binary mixture~\cite{miCasimirBin}, the mean square intensity of concentration fluctuations that is relevant to Casimit forces, can be related to the structure factors discussed in this paper by first considering $S(q,z)$, obtained from the two-points $S(\omega,q,z,z^\prime)$ of Eq.~\eqref{E26}, upon integration over the frequency $\omega$ and substitution of $z=z^\prime$, namely
\begin{equation}
S(q,z)=\frac{1}{2\pi}\int_{-\infty}^\infty d\omega~ S(\omega,q,z,z).
\end{equation}
This quantity is related to the equal-time mean square NE concentration fluctuation at a single spatial point:
\begin{equation}\label{E41}
\langle\delta{c}(\vec{r},t)^2\rangle_\text{NE}=\langle\delta{c}(z)^2\rangle =\frac{1}{4\pi^2}\int_0^\infty 2\pi q~S(q,z)~dq,
\end{equation}
obtained by applying inverse Fourier transforms in the two frequencies and in the two (2D) wave vectors to Eq.~\eqref{E26}. Notice that $\langle\delta{c}(\vec{r},t)^2\rangle_\text{NE}$ is both stationary (it does not depend on $t$) and translationally invariant in the $XY$-plane, \emph{i.e.}, in a direction parallel to the bounding walls (it does not depend on the coordinates $x,y$ of the point $\vec{r}$). It only explicitly depends on the coordinate $z$ of the point $\vec{r}$.

\begin{figure}
\begin{center}
\resizebox{0.90\columnwidth}{!}{\includegraphics{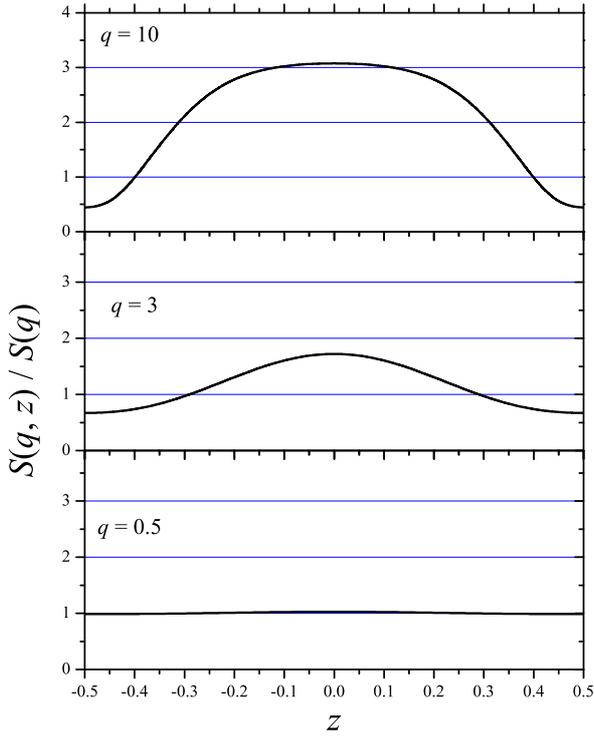}}
\end{center}
\caption{Intensity of the NE concentration fluctuations $S(q,z)$, relative to the value of the structure factor $S(q)$, as a function of $z$ for three wave numbers $q$. In the limit $q\to0$, $S(q,z)$ and $S(q)$ become identical. The deviations increase with increasing $q$, but since the contributions for large $q$ fall off as $1/q^4$, $S(q)$ is a reasonable first approximation in practice.} \label{F1}
\end{figure}

From Eq.~\eqref{E34} for $S(\omega,q,z,z^\prime)$, an explicit expression for $S(q,z)$ can be readily obtained, namely:
\begin{equation}\label{E42}
\begin{split}
S(q,z)& =S(q)+2\sum_{N=0}^\infty\sum_{M=1}^\infty \frac{A_{NM}~\cos(2N\pi z)~\cos(2M\pi z)}{q^2+2N^2\pi^2+2M^2\pi^2} \\
&+2\sum^\infty_{\overset{N,M}{=0}} \frac{B_{NM}~\sin[(2N+1)\pi z]~\sin[(2M+1)\pi z]}{q^2+\dfrac{\pi^2}{2}[(2N+1)^2+(2M+1)^2]}
\end{split}
\end{equation}
with matrices $A_{NM}(q)$ and $B_{NM}(q)$ defined by Eqs.~\eqref{E32} and~\eqref{E33}, respectively, and $S(q)$ being the `experimental' static structure factor given by Eq.~\eqref{E31}. Equation~\eqref{E42} for $S(q,z)$ is a long expression that can only be simplified marginally. Hence, we continue our discussion of $S(q,z)$ graphically.

Figure~\ref{F1} shows the intensity of the NE concentration fluctuations $S(q,z)$ of Eq.~\eqref{E42}, relative to the value of the structure factor $S(q)$ of Eq.~\eqref{E31}, as a function of $z$ for three wave numbers $q$, as indicated. One first observes that, just as for temperature fluctuations in a one-component fluid~\cite{miPRE2014}, $S(q,z)$  has an inhomogeneous spatial distribution that reaches a maximum at mid-layer. However, we stress that, in contrast to the case of NE temperature fluctuations, $S(q,z)$ for the NE concentration fluctuations does not vanish at the walls, $z=\pm\half$. The reason is that to establish a uniform temperature gradient, in experiments the fluid layer is confined between thermally conducting plates~\cite{Mixtures3,SegreEtAlPRE,SegreSengers}, so that the temperature fluctuations, in contrast to the concentration fluctuations, must vanish at the walls of the plates~\cite{Physica,EPJ}.

\begin{figure}
\begin{center}
\resizebox{0.90\columnwidth}{!}{\includegraphics{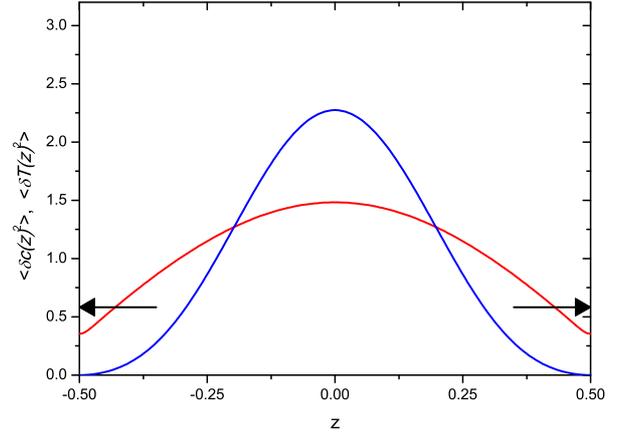}}
\end{center}
\caption{NE $\langle\delta{c}(z)^2\rangle$ as a function of $z$ evaluated from Eqs.~\eqref{E41}-\eqref{E42} (red line) together with NE $\langle\delta{T}(z)^2\rangle$ for rigid boundaries (blue line) as evaluated in previous work~\cite{miPRE2014} for $Pr=6$. For an easy comparison both curves are normalized independently, so that the average value in the layer is, for each case, equal to unity. Note that $\langle\delta{T}(z)^2\rangle$ vanishes at the walls, while $\langle\delta{c}(z)^2\rangle$  does not. The arrows represent the approximation~\eqref{E43} for $\langle\delta{c}^2\rangle_\text{wall}$.} \label{F2}
\end{figure}

Upon substitution of Eq.~\eqref{E42} into Eq.~\eqref{E41} one can obtain an explicit expression for the intensity of NE concentration fluctuations $\langle\delta{c}(z)^2\rangle$. The resulting integrals in $q$ do not admit a compact analytical expression, so that we also discuss this quantity graphically. In Fig.~\ref{F2} the red curve represents the quantity $\langle\delta{c}(z)^2\rangle$ computed by evaluating numerically the $q$-integral in Eq.~\eqref{E41}. It is interesting to compare it with the intensity $\langle\delta{T}(z)^2\rangle$ of the NE temperature fluctuations in a one-component fluid that was investigated in an earlier publication~\cite{miPRE2014} for realistic rigid boundary conditions. Actually, $\langle\delta{T}(z)^2\rangle$ depends on the Prandl number $Pr$~\cite{EPJ}; the blue curve in Fig.~\ref{F2} shows $\langle\delta{T}(z)^2\rangle$ for $Pr = 6$~\cite{miPRE2014}. For easier comparison we normalized the two quantities independently, so that the average value through the layer (area below the curve) is, in each case, equal to unity. Figure~\ref{F2} shows the main conclusion of this section: The different boundary condition (vanishing gradient \emph{vs.} vanishing function) causes different behaviors at the walls, NE temperature fluctuations vanish while NE concentration fluctuations do not. We also emphasize that $\langle\delta{c}(z)^2\rangle$ reaches a non-zero value at the walls with a non-zero slope, although maybe not completely evident in Fig.~\ref{F2} because, due to numerical limitations, only a finite number of modes was added.

To finalize, it is interesting to evaluate the layer-average square of concentration fluctuations, that is the quantity actually relevant for Casimir forces~\cite{miCasimirBin}. Upon substitution of Eq.~\eqref{E31} into Eq.~\eqref{E41} one obtains:
\begin{equation}\label{E43}
\begin{split}
\langle\delta{c}^2\rangle &= \frac{\tilde{F}}{2\pi} \int_0^\infty q~\left[\frac{1}{q^4} + \frac{4(1-\cosh{q})}{q^5(q+\sinh{q})} \right]~dq,\\
&= 3.11 \times 10^{-3}~\tilde{F},
\end{split}
\end{equation}
in terms of dimensionless variables. To revert the result to physical dimensions one has to use Eq.~\eqref{E03} for dimensionless concentration and Eq~\eqref{E19B} for $\tilde{F}$, so that:
\begin{equation}\label{E44}
\langle\delta{c}^2\rangle = 3.11 \times 10^{-3}~\frac{k_\text{B} T}{\rho\nu D} L (\nabla{c})^2.
\end{equation}
We have indicated with arrows the result~\eqref{E44} in Fig.~\ref{F2}, with the same normalization as $\langle\delta{c}(z)^2\rangle$. One observes a small difference between the actual value of $\langle\delta{c}^2\rangle_\text{wall}$ and the value given by Eqs .~\eqref{E43}-\eqref{E44}.

\section{Summary and conclusions\label{S6}}

In this paper we have investigated NE concentration fluctuations in a binary fluid mixture in the absence of gravity. We adopted a realistic setting where a stationary concentration gradient is induced, through the Soret effect, by maintaining the two bounding plates of a fluid layer at different temperatures. No-slip boundary conditions were used for the velocity fluctuations and no-flow for the concentration fluctuations. An approximation of large Lewis and Schmidt numbers was adopted, which means that both temperature and velocity fluctuations decay much more rapidly than the concentration fluctuations, so that the slow dominant mode is a pure concentration mode. This is a good approximation for dense liquid mixtures, while more questionable for rarefied gases.

With all the features described in the previous paragraph, we have been able to obtain exact relatively compact analytical expressions for both the dynamic structure factor, Eq.~\eqref{E30}, and the static structure factor, Eq.~\eqref{E31}, that would be observed in heterodyne low-angle light scattering or shadowgraph experiments. These expressions will be useful for the interpretation of experimental results, in particular in microgravity conditions as well as for computer simulations of NE concentration fluctuations~\cite{CerbinoEtAl}.

We completed the paper studying the total intensity of the NE concentration fluctuations as a function of the distance to the walls. We demonstrated the important result that, due to the different boundary condition, the intensity of concentration fluctuations does not vanish at the walls, in contrast to the case of temperature fluctuations.

\section*{Acknowledgements}

The research at the University of Maryland was supported by the US National Science Foundation under Grant No. DMR-1401449. We also received financial support from the Spanish \emph{Ministerio de Econom\'{\i}a y Competi\-ti\-vi\-dad} by the Research Project FIS2014-58950-C2-2-P.

\section*{Author contribution statement}

All authors contributed equally to the research presented in this paper.

%%%%%%%%%%%%%%%%%%%%%%%%%%%%%%%%%%%%%%%%%%%%%%%%%%%%%%%%%%%%%%%%%%%%%%%%%%%%%%%%%%%%%%%%%%%%%%%%%%%%%%%%%%%%%%%%%%%%%%%%%%%%%%%%%%%%%%%
%\bibliographystyle{spphys}
%\bibliography{ortiz}

%%%%%%%%%%%%%%%%%%%%%%%%%%%%%%%%%%%%%%%%%%%%%%%%%%%%%%%%%%%%%%%%%%%%%%%%%%%%%%%%%%%%%%%%%%%%%%%%%%%%%%%%%%%%%%%%%%%%%%%%%%%%%%%%%%%%%%%
\appendix\onecolumn

\section{Explicit expression for the wall-normal velocity autocorrelation function\label{A2}}
\renewcommand{\theequation}{A\arabic{equation}}

A particular solution of Eq.~\eqref{E23} is:
\begin{equation}\label{EB1}
V_\text{p}(q,z,z^\prime)=\frac{\tilde{F}}{2}\left[|z-z^\prime|~\cosh(q|z-z^\prime|)-\frac{1}{q}\sinh(q|z-z^\prime|)\right].
\end{equation}
The general solution of the homogeneous problem corresponding to Eq.~\eqref{E23} (\emph{i.e.}, by taking the RHS equal to zero) is:
\begin{equation}\label{EB2}
V_\text{h}(q,z,z^\prime)=A_0\cosh{qz} +A_1 \sinh{qz} + A_2 z \cosh{qz} + A_3 z \sinh{qz},
\end{equation}
where the coefficients $A_i$ will be, in general, arbitrary functions of $z^\prime$ and $q$. Adding Eqs.~\eqref{EB1} and~\eqref{EB2}, and imposing the four boundary conditions~\eqref{E25}, the coefficients $A_i(q,z^\prime)$ are uniquely determined. We thus obtain an explicit expression for $V(q,z,z^\prime)$, namely
\begin{multline}\label{E07}
\hspace*{-20pt}V(q,z,z^\prime)=\frac{\tilde{F}}{q^2-\sinh^2{q}}\left\{q^2\left[zz^\prime-\tfrac{1}{4}-\frac{A_2}{q^3}\right]\cosh[q(z-z^\prime)] + A_2  (z-z^\prime) \sinh[q(z-z^\prime)] \right. \\
+ \left. q\sinh{q}\left[zz^\prime+\tfrac{1}{4}+\frac{A_1}{q^2\sinh{q}}\right]\cosh[q(z+z^\prime)] - A_1 (z-z^\prime) \sinh[q(z-z^\prime)] \right\} + V_\text{p}(q,z,z^\prime),
\end{multline}
where
\begin{align}
A_1(q)&=\frac{\sinh{q}+q\cosh{q}}{2},& A_2(q)&=\frac{q+\sinh{q}\cosh{q}}{2}.
\end{align}
We have checked, by numerically evaluating the decay rates $\Gamma_N$ and the modes $W_N(z)$ solving Eq.~\eqref{E13}, for two different values of $q$, that indeed Eq.~\eqref{E07} is equal to the sum of the series~\eqref{E22}.

\section{Explicit expression for the concentration autocorrelation function\label{A3}}
\renewcommand{\theequation}{B\arabic{equation}}

Here, we explicitly present the double trigonometric series for the two-point autocorrelation function $S(\omega,q,z,z^\prime)$ of the NE concentration fluctuations, introduced in Eq.~\eqref{E26}. We first note that there is a slight simplification, since for the function $V(q,z,z^\prime)$ of Eq.~\eqref{E07}:
\begin{equation}
\int_{-1/2}^{1/2} dz\int_{-1/2}^{1/2} dz^\prime~V(q,z,z^\prime)~\cos(2N\pi z)~\sin[(2M+1)\pi z]=0
\end{equation}
for any pair of integers $N,M$. Thus, $S(\omega,q,z,z^\prime)$ reads explicitly:
\begin{multline}\label{E34}
S(\omega,q,z,z^\prime) = \frac{2\tilde{F}}{\omega^2+q^4}\left[\frac{1}{q^2} + \frac{4(1-\cosh{q})}{q^3(q+\sinh{q})} \right]+\sum_{N=1}^\infty \frac{2 A_{N0}(q)~\cos(2N\pi z)}{(-\mathrm{i}\omega+q^2)(\mathrm{i}\omega+q^2+4N^2\pi^2)} \\ +\sum_{N=1}^\infty \frac{2 A_{N0}(q)~\cos(2N\pi z^\prime)}{(\mathrm{i}\omega+q^2)(-\mathrm{i}\omega+q^2+4N^2\pi^2)}
+\sum_{N=1}^\infty\sum_{M=1}^\infty \frac{4A_{NM}(q)~\cos(2N\pi z)~\cos(2M\pi z^\prime)}{(\mathrm{i}\omega+q^2+4N^2\pi^2)(-\mathrm{i}\omega+q^2+4M^2\pi^2)} \\
+\sum_{N=0}^\infty\sum_{M=0}^\infty \frac{4B_{NM}(q)~\sin[(2N+1)\pi z]~\sin[(2M+1)\pi z^\prime]}{[\mathrm{i}\omega+q^2+(2N+1)^2\pi^2][-\mathrm{i}\omega+q^2+(2M+1)^2\pi^2]}.
\end{multline}
The non-zero coefficients of the double trigonometric series~\eqref{E34} are:
\begin{equation}\label{E32}
\begin{split}
A_{NM}(q)&=\int_{-1/2}^{1/2} dz\int_{-1/2}^{1/2} dz^\prime~V(q,z,z^\prime)~\cos(2N\pi z)~\cos(2M\pi z), \\ &=\frac{q^2\tilde{F}~\delta_{NM}}{(q^2+4N^2\pi^2)^2}+\frac{8q^5\tilde{F}(1-\cosh{q})\cos(N\pi)\cos(M\pi)}{(q+\sinh{q})(q^2+4N^2\pi^2)^2(q^2+4M^2\pi^2)^2},
\end{split}
\end{equation}
and
\begin{equation}\label{E33}
\begin{split}
B_{NM}(q)&=\int_{-1/2}^{1/2} dz\int_{-1/2}^{1/2} dz^\prime~V(q,z,z^\prime)~\sin[(2N+1)\pi z]~\sin[(2M+1)\pi z], \\ &=\frac{q^2\tilde{F}~\delta_{NM}}{(q^2+(2N+1)^2\pi^2)^2}+\frac{8q^5\tilde{F}(1+\cosh{q})\cos(N\pi)\cos(M\pi)}{(q-\sinh{q})[q^2+(2N+1)^2\pi^2]^2[q^2+(2M+1)^2\pi^2]^2}.
\end{split}
\end{equation}

\end{document}